\documentstyle[12pt]{article}                    
\newfont{\ffont}{msym10}                          
\newcommand{\beq}{\begin{equation}}               
\newcommand{\eeq}{\end{equation}}                 
\newcommand{\bqry}{\begin{eqnarray}}              
\newcommand{\eqry}{\end{eqnarray}}                
\newcommand{\bqryn}{\begin{eqnarray*}}            
\newcommand{\eqryn}{\end{eqnarray*}}              
\newcommand{\NL}{\nonumber \\}                    
\newcommand{\preprint}[1]{\begin{table}[t]        
            \begin{flushright}                    
            \begin{large}{#1}\end{large}          
            \end{flushright}                      
            \end{table}}                          
\newcommand{\PD}[2]                               
    {\frac{\partial^{#2}}{\partial #1^{#2}}}      
\begin{document}
\preprint{LA-UR-97-3795}
\title{Schwinger's Nonet Mass Formula Revisited}
\author{\\ L. Burakovsky\thanks{E-mail: BURAKOV@PION.LANL.GOV} \
and \ T. Goldman\thanks{E-mail: GOLDMAN@T5.LANL.GOV} \
\\  \\  Theoretical Division, MS B285 \\  Los Alamos National Laboratory \\ 
Los Alamos, NM 87545, USA \\}
\date{ }
\maketitle
\begin{abstract}
We prove the validity of Schwinger's quartic nonet mass formula for a 
general nonet structure with the isoscalar singlet mass being shifted from
its ``ideal'' value. We apply this formula to the problem of the correct 
$q\bar{q}$ assignment for the scalar meson nonet. The results favor
scalar isoscalar singlet and isoscalar octet masses in the vicinity of
1 GeV and 1.45 GeV, respectively. We explain the failure of Schwinger's
formula for the pseudoscalar meson nonet, and suggest a new version of this
formula, modified by the inclusion of the pseudoscalar decay constants, 
which holds with improved accuracy for this nonet. We also rederive the new 
Gell-Mann--Okubo mass formula $3M^2(n\bar{n},I=1)+2M^2(s\bar{s})=4M^2(s\bar{
n})+M^2(n\bar{n},I=0)$ $(n=u,d),$ suggested in our previous publications.
\end{abstract}
\hspace*{0.85in}mesons, scalar mesons 

\section*{ }
Schwinger's original nonet mass formula \cite{Sch} (here the symbol for the
meson stands for its mass)
\beq
\left( 4K^2-3\eta ^2-\pi ^2\right) \left( 3\eta ^{'2}+\pi ^2-4K^2\right) =
8\left( K^2-\pi ^2\right) ^2
\eeq
relates the masses of the isovector $(\pi ),$ isodoublet $(K)$ and isoscalar
mostly octet $(\eta )$ and mostly singlet $(\eta ^{'})$ states of a meson 
nonet.\footnote{Since these designations apply to all spin states, e.g., 
vector mesons will be confusingly labelled as $\eta $'s. We ask the reader to
bear with us in this in the interest of minimizing notation.} It is usually
derived in the following way:

For the ideal Gell-Mann--Okubo type structure of a nonet, the octet-singlet 
mass squared matrix is\footnote{This matrix leads to the Gell-Mann--Okubo 
values for the masses of the pure $n\bar{n}$ and $s\bar{s}$ states $(n=u,d):$
$\pi ^2$ and $2K^2-\pi ^2,$ respectively.}
\beq
M^2=\left( 
\begin{array}{cc}
M^2_{88} & M^2_{89} \\
M^2_{89} & M^2_{99}
\end{array}
\right ) =\frac{1}{3}\left(
\begin{array}{cc}
4K^2-\pi ^2 & -2\sqrt{2}\;(K^2-\pi ^2) \\
-2\sqrt{2}\;(K^2-\pi ^2) & 2K^2+\pi ^2 
\end{array}
\right) .
\eeq
The masses of the physical $\eta $ and $\eta ^{'}$ states are given by 
diagonalizing the $M^2$ matrix, 
\beq
M^2=\left(
\begin{array}{cc}
\eta ^2 & 0 \\
0 & \eta ^{'2}
\end{array}
\right) ,
\eeq
and obtained from the invariance of the trace and determinant under a unitary
transformation, namely a rotation in the flavor space. As follows from (2),(3),
\beq
{\rm Tr}\;\!M^2=2K^2=\eta ^2+\eta ^{'2},
\eeq
\beq
{\rm Det}\;\!M^2=\frac{1}{9}\left[ \left( 4K^2-\pi ^2\right) \left( 2K^2+\pi 
^2\right) -8\left( K^2-\pi ^2\right) ^2\right] =\eta ^2\eta ^{'2}. 
\eeq   
Since, in view of (4),
$$2K^2+\pi ^2=6K^2+\pi ^2-4K^2=3\left( \eta ^2+\eta ^{'2}\right) -\left( 4K^2-
\pi ^2\right), $$
Eq. (5) may be finally cast into the form (1). 

For the ideal structure of a nonet, Eq. (2), the octet-singlet mixing angle,
\beq
\tan 2\theta _{\eta \eta ^{'}}=\frac{2M_{89}^2}{M_{99}^2-M_{88}^2}=2\sqrt{2},
\eeq
corresponds to the ideal value $\theta _{\eta \eta ^{'}}\cong 35.3^o\equiv 
\theta _{id}.$ With this mixing angle, using the relations
\beq
\omega _8=\frac{u\bar{u}+d\bar{d}-2s\bar{s}}{\sqrt{6}},\;\;\;\omega _9=\frac{
u\bar{u}+d\bar{d}+s\bar{s}}{\sqrt{3}},
\eeq
and
$$\omega _\eta =\omega _8\cos \theta _{\eta \eta ^{'}}-\omega _9\sin \theta _{
\eta \eta ^{'}},$$ 
$$\omega _{\eta ^{'}}=\omega _8\sin \theta _{\eta \eta ^{'}}+\omega _9\cos 
\theta _{\eta \eta ^{'}},$$
one finds 
\beq
\omega _\eta =-s\bar{s},\;\;\;
\omega _{\eta ^{'}}=\frac{u\bar{u}+d\bar{d}}{\sqrt{2}}\equiv n\bar{n}.
\eeq
Assuming, as usual, that the relevant matrix elements are equal to the squared
masses of the corresponding states, one obtains from Eq. (2) and the above 
relations \cite{Per}
\bqry
\eta ^2 & = & M^2_{88}\cos ^2\theta _{\eta \eta ^{'}}+M^2_{99}\sin ^2\theta _{
\eta \eta ^{'}}-2M^2_{89}\sin \theta _{\eta \eta ^{'}}\cos \theta _{\eta \eta 
^{'}}\;=\;2K^2-\pi ^2, \\
\eta ^{'2} & = & M^2_{88}\sin ^2\theta _{\eta \eta ^{'}}+M^2_{99}\cos ^2\theta
_{\eta \eta ^{'}}+2M^2_{89}\sin \theta _{\eta \eta ^{'}}\cos \theta _{\eta 
\eta ^{'}}\;=\;\pi ^2,
\eqry
and
\beq
M^2_{\eta \eta ^{'}}\;=\;\left( M^2_{88}-M^2_{99}\right) \sin \theta _{\eta 
\eta ^{'}}\cos \theta _{\eta \eta ^{'}}+M^2_{89}(\cos ^2\theta _{\eta \eta ^{
'}}-\sin ^2\theta _{\eta \eta ^{'}})\;=\;0,
\eeq
in agreement with (3). 

It is easily seen that $\eta $ and $\eta ^{'}$ defined in (9),(10) solve
Schwinger's nonet formula (1). Moreover, (9) and (10) are the unique solution 
of Eqs. (1) and (4). Therefore, Schwinger's nonet mass formula is closely
related to the ideal structure of a nonet: the former has a solution in the
form of the latter, while the latter leads directly to the former, as we have 
demonstrated by Eqs. (2)-(5) above. Indeed, Schwinger's formula is known 
to hold with a high accuracy for, e.g., the vector and tensor multiplets which 
are almost ideally mixed, but it does not hold for the pseudoscalar multiplet.
It is well known that the pseudoscalar (and, probably, scalar) multiplet does 
not have the ideal structure, since the mass of the pseudoscalar isoscalar 
singlet state is shifted up from its ``ideal'' value of $(2K^2+\pi ^2)/3$ (in 
view of (2)), presumably by the instanton-induced 't Hooft interaction 
\cite{tH} which breaks axial U(1) symmetry \cite{symbr,Dmitra,Dmitra2}.
 
Eq. (4) therefore does not hold for the pseudoscalar multiplet, and thus the 
ideal structure of this multiplet is broken. This might be considered as the 
reason for the failure of Schwinger's nonet relation for the pseudoscalar 
mesons. Indeed, with the measured pseudoscalar meson masses, Eq. (1) gives (in
GeV$^2)$ 0.12 on the l.h.s. vs. 0.41 on the r.h.s.. Similarly, according to 
refs. \cite{Dmitra,Klempt}, for the scalar multiplet, the same 
instanton-induced interaction shifts the mass of the scalar isoscalar singlet 
down from the ``ideal'' value provided by Eq. (2) (by the same amount as the 
pseudoscalar isoscalar singlet is increased \cite{Dmitra}), and therefore Eq. 
(4) does not hold for the scalar multiplet, breaking its ideal structure as 
well. Thus, Schwinger's formula may not hold for the scalar mesons either. 
However, this can be neither confirmed nor refuted, as long as the scalar 
meson spectrum remains unknown, except for the scalar isodoublet state 
$K_0^\ast (1430).$ 
  
In this letter we examine a more general structure of a meson nonet than the
ideal one, Eq. (2), viz., the structure with the mass of the isoscalar singlet
shifted from its ``ideal'' value given in (2),\footnote{We assume that the mass
of the isoscalar octet remains in accordance with the Gell-Mann--Okubo mass
formula \cite{GMO} $3M^2_{88}=4K^2-\pi ^2.$}
\beq
M^2_{99}=\frac{1}{3}\left( 2K^2+\pi ^2\right) +A,
\eeq
where $A$ is some constant, with the dimension of mass squared, either positive
or negative (the case $A=0$ corresponds to the ideal structure, Eq. (2)).  
We show that for this structure, Schwinger's nonet formula remains valid.
Hence, this formula should hold for every meson nonet (including the 
pseudoscalar one), and, in particular, since it holds for the scalar nonet, it
may be applied to the problem of the correct $q\bar{q}$ assignment for this 
nonet, which we do below. We show how Eq. (1) must be altered for the 
pseudoscalar mesons, and present a new form of this relation, which holds for 
the pseudoscalar mesons with improved accuracy.
   
We start with the general ``non-ideal'' nonet form which follows from (12):
\beq
M^2=\left( 
\begin{array}{cc}
M^2_{88} & M^2_{89} \\
M^2_{89} & M^2_{99}
\end{array}
\right ) =\frac{1}{3}\left(
\begin{array}{cc}
4K^2-\pi ^2 & -2\sqrt{2}\;(K^2-\pi ^2) \\
-2\sqrt{2}\;(K^2-\pi ^2) & 2K^2+\pi ^2+3A 
\end{array}
\right) .
\eeq
For the pseudoscalar mesons, this structure is reproduced by several different 
QCD-inspired models. For example, the observed mass splitting among the 
pseudoscalar nonet may be induced by the following symmetry breaking terms 
\cite{symbr},
\beq
L^{(0)}_m=\frac{f^2}{4}\left( B\;{\rm Tr}\;\!M\left( U+U^{\dagger }\right) +
\frac{\varepsilon }{6N_c}\left[ \;\!{\rm Tr}\left( \ln U-\ln U^{\dagger }
\right) \right] ^2\;\right) ,
\eeq
with $M$ being the quark mass matrix, 
\beq
M={\rm diag}\;(m_u,\;m_d,\;m_s)=m_s\;{\rm diag}\;(x,y,1),\;\;\;x\equiv \frac{
m_u}{m_s},\;y\equiv \frac{m_d}{m_s},
\eeq
$N_c$ the number of colors, and $B,\varepsilon ={\rm const,}$ in addition to 
the U(3)$_L\times $U(3)$_R$ invariant non-linear Lagrangian 
\beq
L^{(0)}=\frac{f^2}{4}\;{\rm Tr}\;\!\left( \partial _\mu U\partial ^\mu U^
\dagger \right) ,
\eeq
with $$U=\exp (i\pi /f) ,\;\;\;\pi \equiv \lambda _a\pi ^a,\;\;a=0,1,\ldots ,
8,$$ which incorporates the constraints of current algebra for the light 
pseudoscalars $\pi ^a$ \cite{Georgi}. As pointed out in ref. \cite{KM}, chiral
corrections can be important, the kaon mass being half the typical 1 GeV chiral
symmetry breaking scale. Such large corrections are clearly required from 
the study of the octet-singlet mass squared matrix $M^2.$ In the isospin limit
$x=y$ one has \cite{DGH}
\beq
M^2=B\left(
\begin{array}{cc}
\frac{2}{3}(2m_s+m_n) & \frac{2\sqrt{2}}{3}(m_n-m_s) \\
\frac{2\sqrt{2}}{3}(m_n-m_s) & \frac{2}{3}(m_s+2m_n)+\frac{\varepsilon }{BN_c}
\end{array}
\right) ,
\eeq
which reduces, through the Gell-Mann--Oakes-Renner relations (to first order 
in chiral symmetry breaking) \cite{GOR} which we discuss in more detail below,
\bqry
m_\pi ^2 & = & 2\;\!B\;m_n, \NL
m_K^2 & = & B\;(m_s+m_n), \NL
m_\eta ^2 & = & \frac{2}{3}\;\!B\;(2m_s+m_n),\;\;\;m_n\equiv \frac{m_u+m_d}{2},
\eqry
to
\beq
M^2=\frac{1}{3}\left(
\begin{array}{cc}
4K^2-\pi ^2 & -2\sqrt{2}\;(K^2-\pi ^2) \\
-2\sqrt{2}\;(K^2-\pi ^2) & 2K^2+\pi ^2+3A 
\end{array}
\right) ,\;\;\;A\equiv \frac{\varepsilon }{BN_c}.
\eeq
Also, the Nambu--Jona-Lasinio model with the instanton-induced 't Hooft 
interaction, initiated by Hatsuda and Kunihiro \cite{HK} and Bernard {\it et 
al.} \cite{Bern}, and then extensively studied by Dmitrasinovic 
\cite{Dmitra,Dmitra2}, provides shifts of both the pseudoscalar and scalar
isoscalar singlet masses by the same amount but in opposite directions (viz., 
the pseudoscalar isoscalar singlet mass is increased, while the scalar one is
decreased). Thus, for the pseudoscalar mesons, $A$ in (12) may be considered
as the sum of all possible contributions to the shift of the isoscalar singlet
mass (from instanton effects, $1/N_c$-expansion diagrams, gluon annihilation 
diagrams, etc.). 

Let us first calculate the masses of the pure $n\bar{n}$ and $s\bar{s}$ states
(which we now designate as $\eta _n$ and $\eta _s,$ respectively). By rotating 
the $\omega _8$ and $\omega _9$ states in the flavor space by $\theta _{id}$  
and using the relations (9)-(11), one obtains from (12),
\bqry
\eta _s^2 & = & 2K^2-\pi ^2+\frac{A}{3}, \\
\eta _n^2 & = & \pi ^2+\frac{2A}{3}, \\
M^2_{\eta _n\eta _s} & = & -\;\frac{A\sqrt{2}}{3}.
\eqry
It is seen that with $A\neq 0,$ the mass matrix
$$M^2=\left(
\begin{array}{cc}
\eta ^2_s & M^2_{\eta _n\eta _s} \\
M^2_{\eta _n\eta _s} & \eta _n^2  
\end{array}
\right) $$
contains off-diagonal elements, i.e., the $n\bar{n}$ and $s\bar{s}$ states
are not the eigenfunctions of this matrix anymore. Indeed, the octet-singlet
mixing angle obtained from (13) is
\beq 
\tan 2\theta _{\eta \eta ^{'}}=\frac{2M_{89}^2}{M_{99}^2-M_{88}^2}=2\sqrt{2}
\left( 1-\frac{3A}{2(K^2-\pi ^2)}\right) ^{-1},
\eeq
and only with $A=0$ does this reduce to the ideal value (6).\footnote{For the 
pseudoscalar mesons, the parameter $A$ is determined by the trace condition 
(26): $A=\eta ^2+\eta ^{'2}-2K^2\simeq 0.725$ GeV$^2;$ one then obtains from 
(23) the mixing angle $\theta _{\eta \eta ^{'}}\simeq -19^o,$
in agreement with most of experimental data \cite{GK,data}.} Thus, it follows 
clearly that the isoscalar states of a nonet which has the ``non-ideal'' 
structure, Eq. (12), cannot be pure $n\bar{n}$ and $s\bar{s}.$ This is the case
realized for the pseudoscalar and scalar multiplets.

It follows from (20),(21) that 
\beq
2\eta _s^2+3\pi ^2=4K^2+\eta _n^2,
\eeq
which is the new Gell-Mann--Okubo mass formula which we proposed in a previous
paper \cite{GMO_r}. For $A=0$ and $\eta _n=\eta ^{'},$ $\eta _s=\eta $ given 
in (9),(10), Eq. (24) reduces to the Sakurai mass formula \cite{Sakurai}
\beq
2\eta _s^2+\pi ^2+\eta _n^2=4K^2,
\eeq
which may be obtained from (2),(6) \cite{GMO_r,BH}. As is clear from its 
derivation, the formula (24) has more generality than (25), since it holds for
the more general ``non-ideal'' nonet structure, Eq. (12), in contrast to (25) 
which is valid only for the ideal nonet structure alone, Eq. (2).       
 
The eigenvalues of the mass matrix (13), $\eta ^2$ and $\eta ^{'2}$ (the latter
stands for the mostly singlet isoscalar), can be again obtained from the two
matrix invariants:
\beq
{\rm Tr}\;\!M^2=2K^2+A=\eta ^2+\eta ^{'2},
\eeq
\beq
{\rm Det}\;\!M^2=\frac{1}{9}\left[ \left( 4K^2-\pi ^2\right) \left( 2K^2+\pi 
^2+3A\right) -8\left( K^2-\pi ^2\right) ^2\right] =\eta ^2\eta ^{'2}. 
\eeq   
By eliminating $A$ from these two relations, one again arrives at Eq. (1). 
We conclude, therefore, that Schwinger's nonet mass formula is a universal
result of meson spectroscopy, as well as the new Gell-Mann--Okubo mass formula
(24); both hold for a more general (``non-ideal'') structure of a meson nonet. 

We now wish to apply Schwinger's nonet relation (1) to the problem of the 
correct $q\bar{q}$ assignment for the scalar meson nonet. Except for the well
established scalar isodoublet state, $K_0^\ast (1430),$ the masses of the true
scalar states are unknown \cite{pdg}. There are three experimental candidates
for the scalar isovector state, the well-established $a_0(980),$ which is
however interpreted mainly as a $K\bar{K}$ molecule \cite{KKbar}, and the two
states which need experimental confirmation, $a_0(1320)$ and $a_0(1450),$ seen
by GAMS \cite{GAMS} and LASS \cite{LASS}, and the Crystal Barrel \cite{CrBa}, 
respectively, which may well be manifestations of one state having a mass of 
$\sim 1400$ MeV.  
  
We solve the equations (26),(27) for the scalar isoscalar states, using the 
masses of the isodoublet (1429 MeV) and isovector states, for each of the
three isovector candidates discussed above: 983.5 MeV, $1322\pm 30$ MeV, and 
$1450\pm 40$ MeV, respectively. Keeping in mind that the shift of the scalar
isoscalar singlet down from its ``ideal'' value is of the same magnitude but 
opposite in sign to that of the pseudoscalar isoscalar singlet \cite{Dmitra}, 
we use 
\beq
\bar{A}\cong -\left[ \left( \frac{f_\eta }{f_\pi }\right ) ^2\eta ^2+\left( 
\frac{f_{\eta ^{'}}}{f_\pi }\right) ^2\eta ^{'2}-2\left( \frac{f_K}{f_\pi }
\right) ^2K^2\right] \simeq -1\;{\rm GeV}^2,
\eeq
as follows from Eqs. (40),(45) below. 

Our results are shown in Table I. 

We consider the solution for the $a_0(980)$ taken as the scalar isovector 
state to be doubtful, on the basis of the following simple argument: For this 
solution, the mass interval occupied by the scalar nonet turns out to be wider
than 1 GeV, a typical mass scale of light meson spectroscopy. However, we note
that it does afford the most economical solution $[f_0(1700)$ and $f_0^{'}(
500)]$ provided one accepts the arguments of Kisslinger \cite{Kiss}, 
T\"{o}rnqvist \cite{Tor}, and Ishida {\it et al.} \cite{Ishida}, regarding the 
existence of a scalar state in the region of 500 MeV. 

For $a_0=1390\pm 100$ MeV, to accommodate the remaining two candidates for the
scalar isovector state, $a_0(1320)$ and $a_0(1450),$ we have the solution
\beq
f_0=1462\pm 42\;{\rm MeV,}\;\;\;f_0^{'}=971\pm 64\;{\rm MeV,}
\eeq
which strongly favors the $f_0(1500)$ and $f_0(980)$ states of the Particle
Data Group \cite{pdg}, with masses $1503\pm 11$ MeV and $980\pm 10$ MeV, as 
the candidates for the scalar isoscalar octet and singlet states, respectively.

\begin{center}
\begin{tabular}{|c|c|c|} \hline
$a_0,$ MeV & $f_0,$ MeV & $f_0^{'},$ MeV  \\ \hline
 983.5 & 1694 &  463  \\ \hline
  1290 & 1504 &  907  \\ \hline
  1320 & 1485 &  937  \\ \hline
  1350 & 1467 &  966  \\ \hline
  1380 & 1450 &  990  \\ \hline
  1410 & 1436 & 1011  \\ \hline
  1450 & 1423 & 1029  \\ \hline
  1490 & 1419 & 1035  \\ \hline
\end{tabular}
\end{center}
{\bf Table I.} Predictions for the masses of the scalar isoscalar mesons for
different values of the $a_0$ meson mass, according to Eqs. (20),(21). The 
values of $A$ and the $K_0^\ast $ meson mass used in the calculation are $A=1$
GeV$^2,$ $K_0^\ast =1429$ MeV. \\ 

We next explain the inapplicability of Schwinger's formula (1) for the 
pseudoscalar meson nonet, and suggest its proper modification for this nonet.

It was shown by Bramon \cite{Bra} that duality constraints for the set of
scattering processes $\pi \eta \rightarrow \pi \eta ,$ $\pi \eta \rightarrow
\pi \eta ^{'},$ $\pi \eta ^{'}\rightarrow \pi \eta ^{'},$ $\eta K\rightarrow
(\pi ,\eta ,\eta ^{'})K,$ and $\eta \eta \rightarrow \eta \eta ,$ $\eta \eta ^{
'}\rightarrow \eta \eta ^{'},$ $\eta ^{'}\eta ^{'}\rightarrow \eta ^{'}\eta ^{
'},$ provide the value for the $\eta $-$\eta ^{'}$ mixing angle 
\beq
\theta _{\eta \eta ^{'}}=-\arctan \frac{1}{2\sqrt{2}}\approx -19.5^o,
\eeq
in good agreement with experiment \cite{GK,data}. As shown in our previous 
paper \cite{GMO_r}, with this value of the $\eta $-$\eta ^{'}$ mixing angle, 
the formula (24) reduces to
\beq
\eta ^{'2}=4K^2-3\pi ^2,
\eeq
which is satisfied to 1\% with the measured pseudoscalar meson masses. The 
same value (30), on being used in Eq. (23), leads, through (26), to
\beq
A=\eta ^2+\eta ^{'2}-2K^2=3\left( K^2-\pi ^2\right) ,
\eeq
which is satisfied to $\sim 6$\%: (in GeV$^2)$ 0.72 on the l.h.s. vs. 0.68 on 
the r.h.s.. However, Eqs. (31) and (32) give
\beq
\eta =K,
\eeq
which is satisfied only to $\sim 10$\%. The reason for such poor accuracy 
of Eq. (33) (as well as Schwinger's formula (1) for the pseudoscalar nonet
which has the solution $\eta ^{'2}=4K^2-3\pi ^2,$ $\eta ^2=K^2,$ as easily 
checked by using these expressions in Eq. (1)), is the absence of the 
pseudoscalar decay constants from Eqs. (31)-(33), and preceeding relations (19)
and (23). Indeed, the Gell-Mann--Oakes-Renner formulas relate the pseudoscalar
masses and decay constants to the quark masses and condensates \cite{GOR}:
\bqry
\pi ^2f_\pi ^2 & = & -\left[ m_u\langle \bar{u}u\rangle +m_d\langle \bar{d}d
\rangle \right] , \\
\left( K^\pm \right) ^2f_K^2 & = & -\left[ m_u\langle \bar{u}u\rangle +m_s
\langle \bar{s}s\rangle \right] , \\
\left( K^0\right) ^2f_K^2 & = & -\left[ m_d\langle \bar{d}d\rangle +m_s\langle
\bar{s}s\rangle \right] .
\eqry
In the limit of exact nonet symmetry,
\beq
f_\pi =f_K=f_{88}=f_{99}\equiv f,\;\;\;\langle \bar{u}u\rangle =\langle \bar{
d}d\rangle =\langle \bar{s}s\rangle \equiv \langle \bar{q}q\rangle ,
\eeq
one has the mass matrix (17) with $B=-\langle \bar{q}q\rangle /f^2,$ which
further reduces to (19). When the first of the relations (37) is broken, which 
is known to occur in the real world, but the second one remains 
(approximately) satisfied, one will have, from (17) and (34)-(36),
\beq
M^2=\frac{1}{3f^2}\left(
\begin{array}{cc}
4f_K^2K^2-f_\pi ^2\pi ^2 & -2\sqrt{2}\left( f_K^2K^2-f_\pi ^2\pi ^2\right)  \\
-2\sqrt{2}\left( f_K^2K^2-f_\pi ^2\pi ^2\right)  & 2f_K^2K^2+f_\pi ^2\pi ^2+
3f_{99}^2A 
\end{array}
\right) ,
\eeq
$$K^2\equiv \frac{(K^\pm )^2+(K^0)^2}{2}.$$
It then follows from (38), by repeating the considerations which lead to Eqs.
(31)-(33) above, and assuming the validity of Eq. (30), that Eqs. (31)-(33) 
should now be replaced, respectively, by
\beq
\left( \frac{f_{\eta ^{'}}}{f}\right) ^2\!\eta ^{'2}=4\left( \frac{f_K}{f}
\right) ^2\!K^2-3\left( \frac{f_\pi }{f}\right) ^2\!\pi ^2,
\eeq
\beq
\left( \frac{f_{99}}{f}\right) ^2\!A=\left( \frac{f_{\eta }}{f}\right) ^2\!
\eta ^2+\left( \frac{f_{\eta ^{'}}}{f}\right) ^2\!\eta ^{'2}-2\left( \frac{
f_K}{f}\right) ^2\!K^2=3\left[ \left( \frac{f_K}{f}\right) ^2\!K^2-\left( 
\frac{f_\pi }{f}\right) ^2\!\pi ^2\right] ,
\eeq
\beq
\left( \frac{f_\eta }{f}\right) ^2\!\eta ^2=\left( \frac{f_K}{f}\right) ^2\!
K^2.
\eeq  
These relations allow us to determine the ratios $f_\eta /f_\pi ,$ $f_{
\eta ^{'}}/f_\pi ,$ and the values of $f_\eta ,$ $f_{\eta ^{'}},$ using 
$f_K/f_\pi $ and $f_\pi $ as input parameters.

As follows from the relations \cite{pdg} 
\beq
\sqrt{2}f_K=159.8\pm 1.6\;{\rm MeV,}\;\;\;\sqrt{2}f_\pi=130.7\pm 0.3\;{\rm 
MeV},\;\;\;\frac{f_K}{f_\pi }=1.22\pm 0.01
\eeq
and Eqs. (39),(41), 
\beq
\frac{f_{\eta ^{'}}}{f_\pi }=\left[ \left( \frac{f_K}{f_\pi }\right) ^2\left(
\frac{K}{\eta ^{'}}\right) ^2-3\left( \frac{\pi}{\eta ^{'}}\right) ^2\right] ^{
1/2}\!=1.24\pm 0.015,\;\;\;f_{\eta ^{'}}=114.5\pm 1.5\;{\rm MeV,}
\eeq
\beq
\frac{f_\eta }{f_\pi }=\frac{f_K}{f_\pi }\;\frac{K}{\eta }=1.105\pm 0.015,\;\;
\;f_\eta =102\pm 1.5\;{\rm MeV}.
\eeq
Both (43) and (44) are in good agreement with the value provided by the
analysis of experimental data by Gilman and Kauffman \cite{GK}: $f_\eta =f_{
\eta ^{'}}=110\pm 10$ MeV.

Eq. (40), with the relation $f_{99}/f_\pi =1.04\pm 0.04$ obtained from both 
the analysis of experimental $\gamma \gamma $ widths \cite{GK} and chiral 
perturbation theory \cite{DHL}, leads to
\beq
A=3\left[ \left( \frac{f_K}{f_\pi }\right) ^2\left( \frac{f_\pi }{f_{99}}
\right) ^2\!K^2-\left( \frac{f_\pi }{f_{99}}\right) ^2\!\pi ^2\right] =0.97
\pm 0.10\;{\rm GeV}^2\approx 1\;{\rm GeV}^2,
\eeq
which is the value we have used in the analysis of the scalar meson spectrum
above. 

Finally, repeating in the case of the pseudoscalar mesons the analysis of a 
nonet with the ``non-ideal'' structure, Eq. (13), which leads to the 
derivation of Schwinger's formula (1), one arrives at the following 
modification of this formula:
\beq
\left( 4f_K^2K^2-3f_\eta ^2\eta ^2-f_\pi ^2\pi ^2\right) \left( 3f_{\eta ^{
'}}^2\eta ^{'2}+f_\pi ^2\pi ^2-4f_K^2K^2\right) =8\left( f_K^2K^2-f_\pi ^2\pi ^
2\right) ^2,
\eeq
which we refer to as ``Schwinger's nonet mass formula revisited''. With the 
measured pseudoscalar meson masses, and the decay constants (42)-(44), it 
gives (in $10^{-6}$ GeV$^4)$ $$70\pm 7\;\;{\rm on\;the\;l.h.s}\;\;{\rm vs.}\;
\;70\pm 3\;\;{\rm on\;the\;r.h.s},$$ i.e., it holds with a high accuracy for 
the pseudoscalar meson nonet. The uncertainty of Eq. (46) is directly
related to the uncertainties of Eqs. (42)-(44). It is also clear that
\beq
f_{\eta ^{'}}^2\eta ^{'2}=4f_K^2K^2-3f_\pi ^2\pi ^2,\;\;\;f_\eta ^2\eta ^2=f_
K^2K^2,
\eeq
as follows from (39),(41), is a solution to Eq. (46).

It has been suggested in the literature that the pseudoscalar decay constants 
should enter relations like (39)-(41) in the first rather than second power
\cite{first}. As we will show in a separate publication \cite{sep}, the form 
of a relation of the type (39)-(41), containing the pseudoscalar decay 
constants squared, is the only one which remains exact in the first nonleading
order of chiral perturbation theory, while, e.g., the standard Gell-Mann--Okubo
mass formula gets a correction already in this order.

Let us briefly summarize the findings of this work:

i) We have shown that both the new quadratic Gell-Mann--Okubo formula (24) and
Schwinger's quartic formula (1) hold for a more general ``non-ideal'' 
structure of a meson nonet, with the isoscalar singlet mass shifted from its 
``ideal'' value.

ii) We have found a new version of Schwinger's formula for the pseudoscalar 
meson nonet, Eq. (46), which differs from the original relation (1) by the 
inclusion of the pseudoscalar decay constants, and holds with a high accuracy 
for this nonet.

iii) Under the assumption that the $\eta $-$\eta ^{'}$ mixing angle $\approx 
-19.5^o,$ in agreement with experiment, we obtained the pseudoscalar isoscalar
decay constants $f_\eta =102\pm 1.5$ MeV, $f_{\eta ^{'}}=114.5\pm 1.5$ MeV.

iv) We have applied Schwinger's nonet mass formula to the problem of the 
correct $q\bar{q}$ assignment for the scalar meson nonet. Our results favor 
the $f_0(1500)$ and $f_0(980)$ mesons as the candidates for the scalar
isoscalar mostly octet and mostly singlet states, respectively.

\end{document}